\documentstyle[aps,prl,twocolumn,psfig]{revtex}

\def\mpl{M_{Pl}}
\def\mw{M_{EW}}

\def\cc{$\Lambda$}

\makeatletter
\def\vereq#1#2{\lower3pt\vbox{\baselineskip1.5pt \lineskip1.5pt
\ialign{$\m@th#1\hfill##\hfil$\crcr#2\crcr\sim\crcr}}}
\makeatother

\begin{document}

\preprint{LBNL-42967, UCB-PTH-99/07}
\wideabs{
\title{A New Perspective on Cosmic Coincidence 
Problems\cite{acknowledgements}}

\author{Nima Arkani-Hamed, Lawrence J. Hall, Christopher Kolda, 
  Hitoshi Murayama}
\address{Department of Physics, University of California, Berkeley, CA
  94720}
\address{Theoretical Physics Group, Lawrence Berkeley National 
Laboratory, Berkeley, CA 94720}

\maketitle

\begin{abstract}
Cosmological data suggest that we live in an interesting
period in the history of the universe when $\rho_{\Lambda} \sim \rho_{M}
\sim \rho_{R}$.  The occurence of {\it any}\/ epoch with such a ``triple
coincidence'' is puzzling, while the question of why we happen to live
during this special epoch is the ``Why now?'' problem. We introduce a
framework which makes the triple coincidence inevitable; furthermore,
the ``Why now?'' problem is transformed and greatly ameliorated.
The framework assumes that the only relevant mass scales are the electroweak
scale, $M_{EW}$, and the Planck scale, $M_{Pl}$, and requires
$\rho_{\Lambda}^{1/4} \sim \mw^{2}/\mpl$ parametrically. 
Assuming that the true vacuum energy vanishes, we  
present a simple model where a false vacuum energy yields a cosmological 
constant of this form.  
\end{abstract}
\pacs{~}
}
\narrowtext

%

\noindent{\bf 1.}\
Recent observations using high-redshift Type-IA supernovae~\cite{SN} 
indicate a small but non-zero cosmological constant (\cc)
is present in the universe today.
Evidence for a finite \cc\ comes also from the
cosmic age, large scale structure, and probably most convincingly from
the combination of the cosmic microwave background anisotropy and 
cluster dynamics (see \cite{Carroll:2000fy} for recent reviews).  The
data suggest a flat Universe, $\Omega_{\rm total}=1$, with about 70\% of 
the energy density coming from \cc\ and 30\% in the form of non-relativistic
matter. 

\begin{figure}
  \centerline{
  \psfig{file=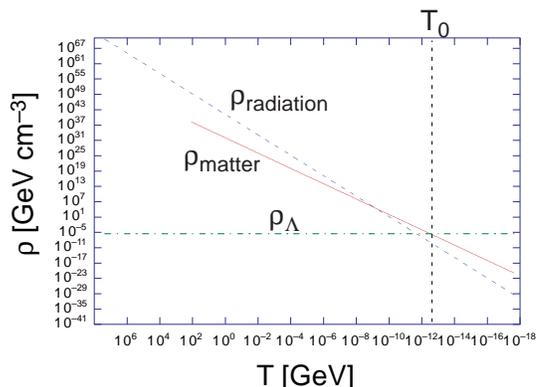,width=0.4\textwidth}}
  \caption{Evolution of the matter, radiation energy densities as a
    function of the radiation temperature and the cosmological
    constant.  Note a near triple coincidence of all three
    around $T_0=2.7$K.  We used $H_0=65~{\rm km}\ {\rm s}^{-1}\ {\rm
      Mpc}^{-1}$, $\Omega_\Lambda=0.7$, $\Omega_M=0.3$.}
  \label{fig:coincidence}
\end{figure}

The tiny energy density of the cosmological constant $\rho_{\Lambda}\sim 
(2 \, $meV$)^4$   
poses the most severe naturalness problem in theoretical physics: 
why is $\rho_{\Lambda} \sim 10^{-120} M_{Pl}^4$? \cite{Weinberg} 
Moreover, the near 
equivalence between the two components in the energy density
raises a serious question: Why should we observe them to be so
nearly equivalent {\it now}\/? While a cosmological constant is by
definition time-independent, the matter energy density is diluted as
$1/R^{3}$ as the Universe expands. Thus, despite evolution of $R$ over
many orders of magnitude, we appear to live in an era during
which the two energy densities are roughly the same (see
Fig.~\ref{fig:coincidence}). This is the ``Why now?'' problem.
If we believe the data, there appears to be no choice but to believe that we
live in a special time in the history of the universe. This 
possibility is thought by many to be sufficiently distasteful 
to warrant disbelieving the data. 

One possible partial remedy to this problem would be to
assume that the ``cosmological constant'' is not a constant at all,
but has always been comparable to
the rest of the energy densities.  Such a possibility is realized in
some models of quintessence which possess ``scaling'' behavior
(see \cite{Albrecht:2000rm} for a recent attempt along this line).
Though such an idea may make the closeness of the effective
\cc\ to the total energy density natural, it does not 
explain the coincidence that the quintessence field becomes
settled with a finite energy density comparable to the matter
energy density just now.

Returning to Fig.~1, we notice {\it two} remarkable features. 
The first is that there is an era in the history of the universe 
where {\it all three} forms of energy, in matter, radiation and $\Lambda$, 
become comparable within a few orders of magnitude. 
The second is that the universe happens to be in this interesting 
era {\it now}. These observations lead to two different questions. 
The first is the ``cosmic triple coincidence'' problem: why should three 
forms of energy density very nearly cross at any point in the evolution of 
the universe? 
The second question is a generalization of the ``Why now?'' problem: why is 
the universe in this era of triple coincidence now? 
The first problem is epoch independent: in principle, at any point in the 
history of the universe the three forms of energy could be measured and 
extrapolated in time, and the triple coincidence could be inferred. 
The existence of a triple coincidence is reminiscent of the unification of 
gauge coupling constants in Grand Unified Theories, 
and suggests an underlying theory in which the coincidence has a simple 
explanation. 

In this letter, we propose a solution to the triple coinicidence problem. 
We begin with one of the central observations of 
%
particle physics, the
special significance played by two mass scales: the reduced Planck scale,
$\mpl\sim 10^{18}$ GeV, and the electroweak scale, $\mw\sim 10^3$ GeV. 
Given these two scales, the
existence of the triple coincidence of matter, radiation and the
cosmological constant can follow in a completely natural way. 
In particular, once
one assumes that \cc\ is order
$\mw^2/\mpl$ (and we will give explicit models in
which this arises), this triple coincidence is inevitable, split only
by ${\cal O}(1)$ coefficients such as $\alpha$ and $\pi$. 
As we will see, this solution to the cosmic coincidence problem also 
allows for a new understanding of the ``Why now?'' problem.

Another apparent ``coincidence'' in cosmology 
arises when one examines the components that
comprise the matter density. Current data favors cold dark matter (CDM) 
as the dominant component of $\Omega_M$, but baryons ($\Omega_B\sim
0.05$) and neutrinos ($\Omega_\nu \sim 0.003$--0.15) are also appreciable.
Having argued for the triple coincidence, we outline how the framework 
can be extended to give a five-fold coincidence: it is natural for 
$\Omega_B$ and $\Omega_\nu$ to also be ${\cal O}(1)$ at the present time. 


\vskip 0.5\baselineskip

\noindent{\bf 2.}\
We begin with the minimal standard model. The physics of electroweak 
symmetry breaking is well-known to suffer from a hierarchy
problem: the energy scale at which the electroweak symmetry
is broken, $\sim {\rm TeV}$, is 15 orders of
magnitude smaller than the Planck scale $10^{18}$~GeV, which we take
to be the fundamental length scale of the universe. Proposed models in
which this hierarchy is explained are substantially more complicated
at the electroweak scale than the standard model itself, with a plethora 
of new particles. Among these, there are typically 
stable particles which can be produced as thermal relics,
such as the lightest supersymmetric particle in models with hidden-sector
supersymmetry breaking~\cite{Goldberg:1983nd}, the lightest messenger
particle in the models of gauge-mediated supersymmetry 
breaking~\cite{Han:1997wn}, and technibaryons in technicolor 
models~\cite{Nussinov:1985xr}, to name a few. 
It is oft-remarked upon 
that the relic density of a stable particle at the
electroweak scale gives the correct order of magnitude for the cold
dark matter component of the Universe, which is believed to dominate
the matter energy density. 
This important result is apparently a numerical coincidence, but in 
our new perspective it will be guaranteed.

It is simple to estimate the relic density 
solely from the fundamental parameters of the theory.
For a stable particle $\chi$ of mass $m$ with electroweak
interactions, its annihilation cross section is given roughly by 
$\sigma \sim 1/m^{2}$. When the temperature falls
below the mass $m$, the expansion rate $H \sim T_f^2/\mpl$
wins over the annihilation rate $\Gamma \sim \sigma n_{\chi}$ and the
abundance $n_\chi/s$ freezes out when $H \sim \Gamma$.  After freeze-out, the
abundance $n_\chi/s$ is constant, so that the  
matter energy density (dominated by $\chi$) is 
\begin{equation}
    \rho_{M} = m_{\chi} n_{\chi} \sim \frac{1}{\sigma} \frac{m^{3}}{\mpl}
        \left( \frac{T}{m}\right)^{3} \sim \frac{\mw^{2} T^{3}}{\mpl},
    \label{eq:rhochi}
\end{equation}
where the factor $(T/m)^3\sim (R_f/R)^3$ is the usual
dilution due to the expansion of the universe after freeze-out.
Given this result, it is easy to estimate the temperature at which
matter-radiation equality occurs, simply by equating $\rho_R = T^4$
with $\rho_M$, and we find
\begin{equation}
T_{eq} \sim \frac{M_{EW}^2}{M_{Pl}}.
\end{equation}

Now, we would like to 
understand the presence of a {\it triple} coincidence, where $\Lambda$ becomes
equal to matter and radiation. If this is not to be a numerical accident, 
it must be that $\Lambda$ itself is determined in terms of the fundamental 
mass scales of the theory as 
\begin{equation}
\rho_{\Lambda} \sim \left(\frac{M_{EW}^2}{M_{Pl}}\right)^4.
\end{equation}
In such a theory, the existence of a triple coincidence between radiation, 
matter and $\Lambda$ is {\it guaranteed} when the universe is at the 
temperature $T_{eq} \sim M^2_{EW}/M_{Pl}$, 
regardless of the numerical value of 
$M_{EW}$. For $M_{EW} \sim 1$ TeV, $T_{eq} \sim 10^{-3}$ eV $\sim$ 10 K.  

We will shortly present models where $\Lambda$ is 
plausibly determined in terms of the fundamental scales in the correct 
combination. However, before discussing this, it is important to note that 
our triple coincidence is  
significantly split by $O(1)$ dimensionless factors.  
A correct version of Eq.~(\ref{eq:rhochi}) is~\cite{KT}:
\begin{equation}
    \rho_{M} = \frac{0.756 (n+1) x_{f}^{n+1}}{g^{1/2} 
    \sigma_{0}\,\mpl} s(T),
    \label{eq:rhochi2}
\end{equation}
where $g$ is the effective number of degrees of freedom at the time of
the freeze-out, $s(T) = \frac{2\pi^{2}}{45} g_{0} T^{3}$ is the entropy
of the Universe, and $x_{f} = m/T_{f} \simeq 20$ is a
parameter which describes the freeze-out temperature $T_{f}$. The
annihilation cross section is thermally averaged as $\langle \sigma v
\rangle = \sigma_{0} x_{f}^{-n}$ with $n=0$ ($n=1$) for $S$-wave
($P$-wave) annihilations.  These dimensionless factors are not very important.  
However, the cross-section $\sigma_0$ 
is suppressed by weak coupling factors so that 
$\sigma_0 \sim \pi \alpha^2/M_{EW}^2$, 
which makes $\rho_M$ bigger by a factor of $\sim 
1/\pi \alpha^{2} \sim 10^{3}$.  This enhancement of the matter energy density
causes the temperature at the matter-radiation equality to become
roughly a factor of $10^3$ bigger, $T_{eq} \sim 1$~eV instead 
of $10^{-3}$~eV. 
The existence of the (approximate) 
triple coincidence is {\it not} a numerical accident, however, since
it is parametrically guaranteed for any values of $\mw$ and $\mpl$.

The following picture emerges from the above considerations.
Based on rough order of magnitude estimates, the triple coincidence
of the matter, radiation energy densities and \cc\ can be 
natural consequences of electroweak physics.
Indeed, the evolution of our Universe shows a near triple coincidence
(Fig.~\ref{fig:coincidence}).  If the coincidence were perfect, the
Universe would be rather boring: 
before $T_{eq} \sim \mw^2/\mpl$
it is radiation dominated and structure cannot form; after $T_{eq}$
the Universe starts to inflate, leaving a completely empty
Universe.  However the $O(10^{3})$ enhancement in $\rho_{M}$ causes
matter-radiation equality to occur
before matter-\cc\ equality. Thus, as can be seen in
Fig.~\ref{fig:coincidence}, there is a small triangle
during which density fluctuations can grow, resulting in 
structure. If the cosmological constant is truly a constant, then this
interesting period does not last long; the matter energy density soon 
dilutes to the level of \cc\ and this interesting period ends. On the other 
hand, the apparent cosmological constant could be due to an energy density 
in a quintessence field $\phi (\sim \mpl)$ of 
mass $m_{\phi}^2 \sim M_{EW}^8/M_{Pl}^6$, which 
has been prevented from falling by the Hubble friction until the triple 
coincidence era. In this case, $\phi$ will fall to the minimum
of its potential a few e-foldings into the coincidence era, and the universe
will return to being matter-dominated.  
 

\vskip 0.5\baselineskip

\noindent{\bf 3.}\
We would now like to demonstrate that it is natural to expect \cc\ 
of order $(\mw^2/\mpl)^4$. We will start with a
particularly simple model which uses false vacua as the origin of the 
\cc\ \cite{carlson}. We
do not insist that this model should be the true origin of \cc, 
but we find it extremely encouraging that such a
simple model serves the purpose.  We assume supersymmetry, which is
broken at the TeV scale by an order parameter chiral superfield
$\langle S \rangle = \theta^{2} \mw^2$, so that the electroweak
symmetry breaking is indeed the direct consequence of the
supersymmetry breaking.  We also assume that there is a ``hidden
sector,'' which couples to the supersymmetry breaking only by
Planck-scale suppressed operators.  This is reminiscent of the
conventional hidden sector supersymmetry breaking models in
supergravity except that we now reverse the roles of the hidden and 
the observable sectors. In the hidden sector, we have a
supersymmetric QCD with $SU(n_{c})$ gauge group (or any other
non-abelian gauge theory)
with $n_{f}$ flavors $Q+\tilde{Q}$, and assume that the
beta function $3n_{c}-n_{f}$ is somewhat small.  Once supersymmetry is
broken in the observable sector by the order parameter $S$, it can
generate masses for the hidden sector quarks by
\begin{equation}
    \int d^{4}\theta \frac{S^{*}}{\mpl} \tilde{Q}Q
    = \int d^{2} \theta \frac{\mw^2}{\mpl} \tilde{Q}Q.
    \label{eq:GM}
\end{equation}
Therefore, the masses of quarks and squarks are of the order of $m_{Q} \sim
\mw^2/\mpl$.  Similarly, the gluino $\lambda$ of the
$SU(n_{c})$ gauge group also acquires a mass via the operator
\begin{equation}
    \int d^{2} \theta \frac{S}{\mpl} W_{\alpha} W^{\alpha}
    = \frac{\mw^2}{\mpl} \lambda \lambda,
    \label{eq:gluino}
\end{equation}
and again the mass is of the order of $m_{\lambda}/g^{2} \sim
\mw^2/\mpl$.  For the simplicity of the analysis, we
take the gluino mass somewhat smaller than the others.  In the absence
of the gluino mass, the decoupling of the quarks generates the
low-energy dynamical scale:
\begin{equation}
    \Lambda_{\rm low}^{3n_{c}} = \Lambda_{\rm high}^{3n_{c}-n_{f}} 
    m_{Q}^{n_{f}}.
    \label{eq:QCDscale}
\end{equation}
Note that the low-energy dynamical scale is determined essentially by
$m_{Q}$ if $3n_{c}-n_{f} \ll n_{f}$.  It is basically that the gauge
coupling constant evolves slowly above $m_Q$, while it grows
quickly below $m_Q$.  

The low-energy theory, supersymmetric pure
Yang--Mills, develops a gluino condensate $\langle \lambda
\lambda \rangle$ with superpotential
\begin{equation}
    \int d^{2}\theta\, \Lambda_{\rm low}^{3} e^{2\pi i k/n_{c}}.
    \label{eq:Weff}
\end{equation}
Here, the integer $k=1, \cdots, n_{c}$ parameterizes the degenerate
vacua consistent with the Witten index $n_{c}$.  The discrete
$R$-symmetry $Z_{2n_{c}}$ is broken spontaneously down to $Z_{2}$ by
the gluino condensate.  In the presence of the (small) gluino mass, we
replace the gauge coupling constant as $8\pi^{2}/g^{2} \rightarrow
8\pi^{2}(1 + \theta^{2} m_{\lambda})/g^{2}$ which breaks the
$R$-symmetry explicitly to $Z_{2}$.  The vacuum energy can be
calculated exactly to lowest order in $m_{\lambda}$:
\begin{equation}
    V_{k} = \frac{16\pi^{2}}{g^{2}} |m_{\lambda} \Lambda_{\rm low}^{3}|
        \left(c - \cos \frac{2\pi k+\Theta}{n_{c}}\right),
    \label{eq:Vk}
\end{equation}
where $\Theta$ is the $\Theta$-parameter of $SU(n_c)$.
We assume the ground state energy is tuned to zero by choosing the
constant $c$ appropriately. However the system 
can drop into any of the states labeled by $k$ and in general has the 
vacuum energy of the order of $m_{\lambda} \Lambda_{\rm low}^{3} \sim 
(\mw^2/\mpl)^{4}$.  Although in this model we assumed 
low-energy supersymmetry breaking, a simple variation
can be extended to the case of gravity-mediated  supersymmetry 
breaking \cite{TBD}. If the $\Theta$-parameter 
is dynamical, {\it i.e.}, $\Theta=a/\mpl$ for an axion-like field $a$, 
then $a$ has the properties of a quintessence
field, beginning a slow roll down its potential only recently and
contributing $\rho_a\sim (\mw^2/\mpl)^4$~\cite{TBD}.


In the above model, we assumed that an unknown mechanism cancelled the 
vacuum energy at the global minimum of the potential. It may be that 
the true mechanism for cancelling the cosmological constant
has the feature that only {\it part} of the vacuum energy is cancelled. 
Indeed, \cite{ccp} propose to eliminate the purely Standard
Model loop contributions to the cosmological constant, by embedding the 
Standard Model fields on a 3-brane in large extra dimensions. 
In these models, the effective four dimensional cosmological 
constant is a power series 
expansion in powers of the Standard Model vacuum energy $V_{SM} 
\sim M_{EW}^4$ of the form 
\begin{equation}
    \rho_{\Lambda} = \sum_{n=0}^{\infty} c_n 
    \frac{V_{SM}^{n+1}}{M_*^{4n}}
\end{equation}
where $M_*$ is the higher dimensional Planck scale related to the 
4D Planck scale as $M_{Pl}^2 = M_*^6/V_{SM}$, and the 
coefficient $c_0 = 0$ naturally. Unfortunately, the terms with 
$n=1,2$ are still too large, but $n=3$ gives the perfect 
combination $\rho_{\Lambda,4} \sim (M_{EW}^2/M_{Pl})^4$. 
These models demonstrate that an incomplete cancellation
mechanism for truly solving the cosmological constant problem may yield a 
$\Lambda$ appropriately determined by $M_{EW}$ and $M_{Pl}$. 

\vskip 0.5\baselineskip

\noindent{\bf 4.}\ 
What about neutrino and baryon energy densities?
The fact that the neutrino energy density is
comparable to the rest has a natural explanation if the neutrino
mass arises from the seesaw mechanism.  In terms of the two
fundamental scales $\mw$ and $\mpl$, the seesaw mechanism gives $m_\nu
\sim \mw^2/\mpl$.  Then we find
\begin{equation}
  \rho_\nu \sim m_\nu T_0^3 \sim \frac{M_{EW}^2}{\mpl} T_0^3 
  \sim \frac{\mw^8}{\mpl^4},
\end{equation}
again of the same order of magnitude as $\rho_M$, $\rho_R$ and
$\rho_{\Lambda}$.  One can also find models of baryogenesis
where the baryon asymmetry is generated naturally as $n_{B}/s \sim 
10^{5}\, \mw/\mpl$ where $10^{5} \sim (2\pi/\alpha)^{2}$ \cite{TBD}.  Then 
the baryon energy density is 
\begin{equation}
  \rho_B \sim 10^5 \frac{\mw}{\mpl} T_0^3 m_p
  \sim 10^5 \frac{m_p}{\mw} \frac{\mw^8}{\mpl^4},
\end{equation}
which is also parametrically the same combination of $\mw$ and $\mpl$
if one regards $m_p/\mw \sim 10^{-3}$ as an $O(1)$ coefficient.

\vskip 0.5\baselineskip

\noindent{\bf 5.}\ 
Given our solution to the triple coincidence problem, we gain a
significant insight into the ``Why now?'' problem.  For this purpose, we
need to define the problem more precisely.  One aspect of the ``Why
now?''  problem is that the time when structure starts to form $t_{\rm
  str}$ and the time of matter-$\Lambda$ equality $t_\Lambda$ are close to
each other.  Within our solution to the triple coincidence problem, 
this means
\begin{equation}
  t_{\rm str} \sim (\pi\alpha)^{2} \frac{\mpl^{3}}{\mw^{4}} 
        \left( \frac{\delta \rho}{\rho} \right)_{p}^{-3/2}
        \sim t_\Lambda \sim \frac{\mpl^{3}}{\mw^{4}} .
\end{equation}
Here, $(\delta \rho/\rho)_{p} \sim 10^{-5}$ is the primordial density fluctuation
generated during inflation.  Note that
both time scales have the same parametric dependences on $\mpl$ and $\mw$,
which makes this near equality natural.  This is essentially the same as the
statement that the triple coincidence is natural in our framework.
In fact, the equality is 
surprisingly 
good, given that our solution by itself does not fix unknown $O(1)$
coefficients.  

Another aspect of the ``Why now?'' problem is why we do not live at a time 
far beyond $t_{\rm str}$.  This 
is an old mystery present even if the cosmological constant were zero 
and not 
our main concern here.  Nonetheless it is interesting to discuss it
briefly.  Once structure forms, life as we know it is limited to the 
time before which all the available hydrogen has been burnt up in stars.
The time scale for this is given by
\begin{equation}
  t_{\rm burn} \sim 
  10^{2} \frac{\pi\alpha^{2}8\pi\mpl^{2}}{m_{p} m_{e}^{2}}
  \sim \frac{\mpl^{2}}{\mw^3} \frac{\mw}{m_p h_e^2}
\end{equation}
where $h_e$ is the electron Yukawa coupling.  It is an old puzzle why
$t_{\rm burn} \sim t_{\rm str}$.  Having expressed $t_{\rm str}$ in
terms of the fundamental energy scales, the dependence on $\mpl$ and
$\mw$ suggests that $t_{\rm burn}$ is in general much shorter than
$t_{\rm str}$.  This would actually strengthen our solution to the
``Why now?'' problem since it would leave little room between structure
formation at $t_{\rm str}$ and the time when all nuclear fuel is used up at
$t_{\rm str} + t_{\rm burn}$; then $t_{\rm str} \sim t_\Lambda$ would 
be enough to solve the puzzle.  For a wide range of dimensionless
parameters $m_p/\mw$ and $h_e$, this near equality remains true.  In
reality, $h_e$ is so small that $t_{\rm burn} \sim 100 t_{\rm str}$.
The fact that $t_{\rm str}$ and $t_{\rm burn}$ are nearly equal in orders of
magnitude is a purely numerical coincidence beyond the scope of 
our current understanding. 
Note, however, that the only requirement for solving the ``Why now?''
problem is that the two dimensionless quantities $m_p/\mw$ and $h_e$
are not too small.

If $m_p/\mw$ and/or $h_e$ were smaller than observed,
stars could still exist much later 
than $t_{\rm str} \sim t_{\Lambda}$.  However, it is intriguing to note 
that cosmology cannot be done long after the  
matter-$\Lambda$ equality.  Structures at cosmological distances go 
beyond the horizon, the cosmic microwave background is 
exponentially redshifted and nuclei become highly processed.  Even if 
some life-forms remained, they might not be able to perform 
cosmological observations to discover the Big Bang.



\vskip 0.5\baselineskip

\noindent{\bf 6.}\
In summary, we have introduced a new perspective on the cosmic coincidence
problems.  We begin with the assumption that there are
only two fundamental energy scales
in the problem, the Planck scale $\mpl\sim 10^{18}$~GeV and
the electroweak scale $\mw \sim 1$~TeV. Electroweak scale
physics produces a cold dark matter relic, while the cosmological
constant is also tied to the electroweak physics as $\rho_{\Lambda}^{1/4}
\sim \mw^2/\mpl$.  Then we find that a triple
coincidence among the matter, radiation, and \cc\ 
energy densities is a necessary consequence.  
Putting $O(1)$
coefficients back into the discussion, the matter energy density is 
enhanced by a factor of $1/\pi\alpha^{2} \sim 10^{3}$, and
a cosmological window opens in the
evolution of the Universe between the matter-radiation equality and
\cc-dominance during which structure forms.
We presented a simple model in which
the cosmological constant is indeed generated at $\rho_{\Lambda}^{1/4}
\sim \mw^2/\mpl$.  Moreover, we pointed out that the 
coincidence of the neutrino as well as the baryon energy density can 
also be simply understood.
This framework allows us to understand why the time scale for 
structure formation and matter-\cc\ equality are comparable in orders 
of magnitude, solving the aspect of the ``Why now?'' problem 
involving the cosmological constant.

%

\end{document}